\documentclass[]{aastex631}

\newcommand{\Chandra}{\mbox {\em Chandra}}
\newcommand{\um}{$\mu$m}
\newcommand{\hii}{\mbox{\ion{H}{2}~}}
\renewcommand{\deg}{$^\circ$}
\newcommand{\degree}{^{\circ}}

\newcommand{\kms}{km~s$^{-1}$}
\newcommand{\peryr}{yr$^{-1}$}

\newcommand{\anchorparen}[2]{\href{#1}{#2} (\url{#1})}
\submitjournal{AAS Journals}

\begin{document}

\title{Structure and Large-Scale Kinematics of Young Stellar Populations in the NGC 6357 and NGC 6334 Giant Molecular Cloud Complex}

\correspondingauthor{Matthew S. Povich}
\email{mspovich@cpp.edu}

\author[0000-0001-9062-3583]{Matthew S. Povich}
\affiliation{Department of Physics and Astronomy,
California State Polytechnic University,
3801 West Temple Avenue,
Pomona, CA 91768, USA}

\author[0000-0001-8081-9152]{Leisa K. Townsley}
\altaffiliation{Deceased.}
\affiliation{Department of Astronomy and Astrophysics,
The Pennsylvania State University, 525 Davey Laboratory, 
University Park, PA  16802, USA}

\author[0000-0002-7872-2025]{Patrick S. Broos}
\affiliation{Department of Astronomy and Astrophysics,
The Pennsylvania State University, 525 Davey Laboratory, 
University Park, PA  16802, USA}

\author{Aldair E. Bonilla}
\affiliation{Department of Physics and Astronomy,
California State Polytechnic University,
3801 West Temple Avenue,
Pomona, CA 91768, USA}

\author{Giaky Nguyen}
\affiliation{Department of Physics and Astronomy,
California State Polytechnic University,
3801 West Temple Avenue,
Pomona, CA 91768, USA}

\author{Carly Soos}
\affiliation{Department of Physics and Astronomy,
California State Polytechnic University,
3801 West Temple Avenue,
Pomona, CA 91768, USA}

\author[0000-0002-0631-7514]{Michael A. Kuhn}
\affiliation{Centre for Astrophysics Research, Department of Physics, Astronomy and Mathematics, University of Hertfordshire, Hatfield, AL10 9UW, UK}

\author[0009-0006-3386-6160]{Simran S. Singh}
\affiliation{Centre for Astrophysics Research, Department of Physics, Astronomy and Mathematics, University of Hertfordshire, Hatfield, AL10 9UW, UK}

\author[0000-0002-7371-5416]{Gordon P. Garmire}
\affiliation{Huntingdon Institute for X-ray Astronomy, LLC, 10677 Franks Rd, Huntingdon, PA 16652, USA}



\begin{abstract}
We map the three-dimensional structure and large-scale kinematics of the young stellar populations in the G352 giant molecular cloud (GMC) complex. In radio and infrared images, G352 appears as long filament extending ${\sim}3\degree$ (${\sim}150$~pc) parallel to the Galactic midplane. It connects the NGC 6357 and NGC 6334 giant \hii regions and the GM1-24 compact \hii~region.
We identify 1727 stellar members of G352 via matching large catalogs of Chandra X-ray point sources and Spitzer mid-infrared excess sources to the Gaia DR3 astrometric catalog. 
Our catalog of 11,470 X-ray point sources ranks among the three largest contiguous X-ray survey datasets ever assembled for a massive star-forming complex. 
We revise the mean heliocentric distance of G352 to $1670\pm 80$~pc, with the median parallaxes of seven constituent groups exhibiting a trend toward increasing distance with decreasing Galactic longitude. We identify two foreground stellar groups superimposed on NGC 6357 that may belong to the Sag OB4 association. The three massive clusters in NGC 6357 exhibit peculiar velocities that trail Galactic circular motion by ${\sim}8$~\kms, while the stars associated with NGC 6334 are more consistent with a circular orbit. GM1-24 has a distinct proper motion and smaller parallax compared to NGC 6334. The steep pitch angle of the GMC filament into the sky appears inconsistent with a spiral arm. The various stellar groups are not gravitationally bound to each other, making G352 a proto-OB association.
\end{abstract}



\section{Introduction} \label{sec:intro}

The spectacular ``mini-starburst'' regions NGC 6357 (sometimes known as the Lobster Nebula or War and Peace Nebula) and NGC 6334 (the Cat's Paw Nebula) illuminate a filamentary giant molecular cloud (GMC) complex extending ${>}150$~pc along the inner Galactic plane \citep{russeil16,russeil17,fukui18a}. 
 Gaia DR2 parallaxes for various constituent stellar populations associated with NGC 6357 and NGC 6334 placed G352 at $d=1780$~pc \citet{kuhn19,russeil20}. G352 is one of the closest massive star-forming complexes to the Sun, with total luminosity $L_{\rm bol}>7\times 10^6~L_{\odot}$ and ionizing photon production rate $Q_0 \approx 3.7\times 10^{50}$~s$^{-1}$ \citep{BP18}.
  
 The most famous and well-studied star cluster in G352 is the optically-revealed Pismis 24 \citep{pismus59} in NGC 6357, which contains some of the most massive stars known in the Galaxy \citep[spectral types O3.5 III--I;][]{walborn02}.
Two other rich clusters in NGC 6357 suffer greater extinction than Pi 24 but also host OB stars, including another O3 star candidate \citep{MYStIX_OBc,macla20}. In comparison to NGC 6357, the stellar populations in NGC 6334 are even more obscured and embedded.  
A high level of ongoing star formation activity in NGC 6334 is evidenced by its rich populations of young stellar objects \citep[YSOs;][]{MIRES,willis13} and ultra-compact \hii~regions \citep{brogan16}. A mismatch between the nebular luminosity and the total bolometric luminosity of the known and candidate OB stellar population indicates that half of the massive stars in NGC 6334 have yet to be directly identified \citep{BP18}. Extremely bright nebular emission at visible through IR wavelengths hampers detection of point sources in both giant \hii~regions.

G352 has not been as well-studied historically as the comparably large, luminous, and nearby Galactic star-forming regions Cygnus-X \citep{CygOB2} and Carina \citep{CCCP_Intro}.
This may be changing. NGC 6357 has been the target of multiple guest observer programs with the James Webb Space telescope, as a laboratory for the effects of external irradiation on the evolution of protoplanetary disks \citep{XUE1,XUE10,XUE_survey}. The upcoming Galactic Plane Survey with the Nancy Grace Roman Observatory has selected the entire G352 field encompassing NGC 6357 to NGC 6334 as a time-domain target for repeated visits in multiple near-IR filters \citep{RomanGPSdef}. 

G352 also occupies a pivotal location for mapping Galactic structure beyond the immediate solar neighborhood. While the GMC complex has long been considered part of the Sagittarius--Carina spiral arm \citep{reid19,russeil20}, recent studies have cast doubt on the location or even the existence of this structure.
\citet{Sagspur21} mapped star formation across the first Galactic quadrant, using Gaia DR2 parallaxes and proper motions of dozens of YSO groups, many of them associated with prominent \hii regions including the Lagoon Nebula, M17, and the Eagle Nebula. The  ``Sagittarius spur'' structure connecting the first-quadrant star-forming regions between $d=1$ and 2 kpc has too high of pitch angle to be a true spiral arm, a finding corroborated by 3D dust mapping \citep{green19}.
Based on its longitude and distance, G352 could mark the reappearance of the Sagittarius--Carina Arm in the 4th quadrant. However, the recent southern extension of the dust maps by \citet{Z25} showed an absence of dust along the location of the predicted spiral arm south of G352.

This paper is motivated by a desire to help guide future studies of NGC 6357 and NGC 6334. We have two main goals: (1) To provide a more detailed map giving the locations, distances, and kinematics of various constituent young stellar populations; and (2) To complete and publish in its entirety our X-ray survey data covering the full length of G352. %, from NGC 6357 to GM 1-24 \citep{GM77}.

Most of our X-ray data were previously published as part of the Massive star-forming regions Omnibus X-ray Catalog \citep[MOXC;][]{MOXC,MOXC2,MOXC3}.
In Section~\ref{sec:data} we add new X-ray data to the MOXC archive, introduce the other multiwavelength source catalogs used in this work, and describe our process for combining and refining these lists.
Our astrometric selection of probable G352 complex members from our initial source list is described in Section~\ref{sec:selection}. Our results, discussion, and conclusions are presented in Sections~\ref{sec:results} and \ref{sec:conclusions}.

\section{Multiwavelength Point-Source Catalogs}\label{sec:data}

\subsection{MOXC Joined by a New \Chandra/ACIS Source Catalog}

\begin{figure}[htb]
\centering
\plotone{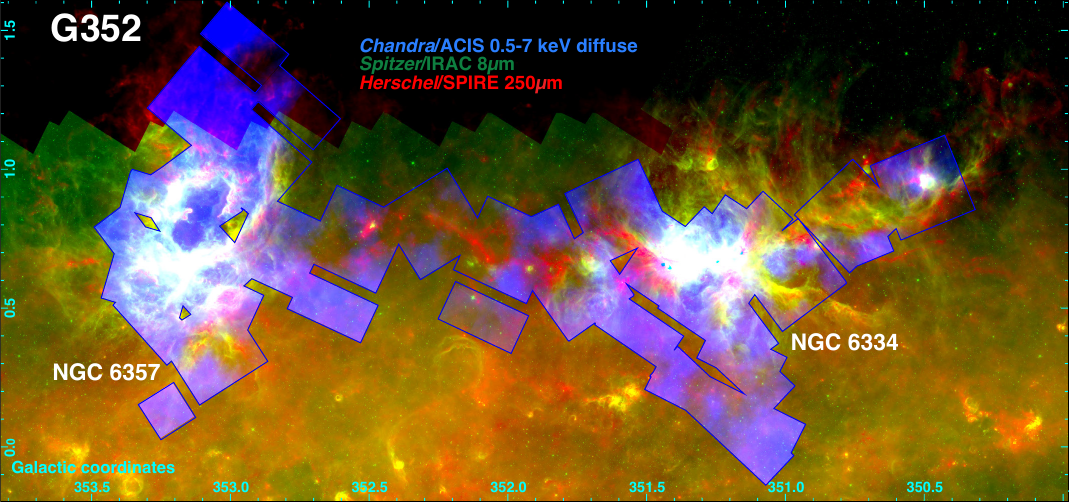}
\caption{Wide-field ACIS mosaic of the G352 giant molecular filament connecting NGC~6357 and NGC~6334.  Total-band (0.5--7~keV) ACIS diffuse emission (blue) is shown with Spitzer MIR (green) and Herschel FIR (red) mosaics. To highlight diffuse X-ray emission, 11,470 X-ray point sources were excised from the ACIS mosaic.
\label{g352fullfield.fig}
}
\end{figure}

The acquisition and processing of new \Chandra/ACIS data that bridge the IRDC connecting NGC~6537 and NGC~6334 provides a primary motivation for this study. The resulting X-ray mosaic spans $3\degree$ parallel to the Galactic plane and contains 11,470 detected point sources (Figure~\ref{g352fullfield.fig}), placing it among the three largest contiguous X-ray datasets surveying GMC complexes (Table~\ref{tab:Xraysurveys}).

%---the previous two were the Chandra Carina Complex Project \citep[CCCP, ${\sim}14,000$ point sources;][]{CCCP_Intro} and the Cygnus OB2 Legacy Survey \citep[${\sim}8,000$ point sources;][]{CygOB2}.

\begin{deluxetable}{lcccccrc}\label{tab:Xraysurveys}
\centering 
\tablecaption{Three Largest Chandra Surveys of Galactic GMC Complexes}
\tablehead{
\colhead{Survey} &
\colhead{d} &
\colhead{} &
\colhead{Angular} &
\colhead{Nom. Exp.} &
\colhead{Tot. Exp.} &
\colhead{\# point} &
\colhead{} \\
\colhead{Region} &
\colhead{(kpc)} &
\colhead{\# ObsIDs} &
\colhead{Dimensions} &
\colhead{(ks)} &
\colhead{(Ms)} &
\colhead{sources} &
\colhead{Referenc(es)}
}
\startdata
CCCP & 2.3 & 38 & $1.4\degree\times1.2\degree$ & 60 & 1.2 & 14,000 & \citet{CCCP_Intro} \\
CygOB2 & 1.4 & 41 & $1\degree\times1\degree$ & 100 & 1.2 & 8,000 & \citet{CygOB2} \\
G352 & 1.7 & 24 & $3\degree\times 0.5\degree$ & 40 & 0.8 & 11,500 & \citet{MOXC2,MOXC3}, this work
\enddata
\end{deluxetable}
%
%\todo{Edit this to reflect what Leisa and Pat actually did! 2 old exposures were discarded due to high background; let's not bother including those in our observing log, just mention in the text!}
We obtained 5 new ACIS pointings (one of which was divided over two ObsIDs) and extracted 1,560 point sources (=787 from the three GO pointings + 773 from the two GTO pointings; 
see Appendix \ref{sec:newdata} for details). 
The data were processed using the MOXC strategies and pipeline procedures, previously employed to produce source lists covering NGC 6334 \citep[4674 point sources;][]{MOXC2} and NGC 6357 \citep[5269 point sources;][]{MOXC3}.
The new data overlap with the MOXC mosaics in a few areas at far off-axis angles, and only 33 sources are found in common with the previously published MOXC catalogs. 
%NGC 6357 - 5,269 sources \citep[MOXC3;][]{MOXC3}
%NGC 6334 - 4,674 sources \citep[MOXC2;][]{MOXC2} 
For the convenience of the community, we include a unified X-ray point source catalog spanning the entire G352 complex, published as an electronic table accompanying this article (see Appendix~\ref{sec:newdata}). This catalog has been astrometrically registered to Gaia ICRS coordinates using the same procedure described by \citet{MOXC2} for registration to 2MASS.

\subsection{SPICY Catalog}
The Spitzer/IRAC Candidate YSO (SPICY) catalog contains ${\sim}120,000$ objects selected based on detectable mid-infrared (MIR) excess emission due to the presence of dusty circumstellar disks and/or infalling envelopes \citep{SPICY}. The SPICY footprint covering G352 generally follows the 8~\um\ mosaic boundaries visible in Figure~\ref{g352fullfield.fig}, hence it includes all of the X-ray mosaic except for the portion of NGC 6357 at $b>1.0\degree$.
We selected candidate YSOs in the vicinity of G352 using a rectangular area superscribing the  ACIS mosaic boundary in Galactic coordinates. The box defined as $353.469\degree > l > 350.335\degree$ and $-0.057\degree < b < 1.517\degree$ contains 2,336 SPICY objects. Using a $1\arcsec$ sky matching radius, we identified 304 potential common objects between SPICY and our X-ray point-source catalog, for a combined X-ray and MIR-excess selected list of 13,502 sources.

\subsection{Gaia DR3}
The same $3.2\degree\times 1.5\degree$ box was used to query the Gaia DR3 catalog \citep{GaiaDR3}. Unlike SPICY or MOXC, Gaia is a visible-light all-sky survey \citep{gaia_collab_mission} that completely covers the target area for our study, and the query returned 306,591 sources. Because of the very high extinction produced by G352 itself, the Gaia source density is significantly lower near the cloud ridge, where the \Chandra\ observations were targeted and the SPICY catalog sources tend to congregate. A $1\arcsec$ sky coordinate match identified 3,218 Gaia DR3 sources with counterparts in our unified X-ray catalog, SPICY, or both.

We further refined our sample by selecting only those Gaia sources considered to pass a conservative single-star astrometric solution by satisfying RUWE $<1.4$ \citep{RUWE24}. This kept 2,519 candidate young stars in our final table: 1312 in the vicinity of NGC 6357 ($l > 352.7\degree$), 1087 in NGC 6334 ($l < 351.8\degree$), and 120 in the IRDC bridge.

%%\noindent {\tt\string\documentclass[twocolumn]\{aastex631\}}. \\

%%\noindent Note that in the two column style figures and tables will only
%%span one column unless specifically ordered across both with the ``*'' flag,
%%e.g. \\
%%
%%\noindent{\tt\string\begin\{figure*\}} ... {\tt\string\end\{figure*\}}, \\
%%\noindent{\tt\string\begin\{table*\}} ... {\tt\string\end\{table*\}}, and \\
%%\noindent{\tt\string\begin\{deluxetable*\}} ... {\tt\string\end\{deluxetable*\}}. \\

%%Full details on how to create the different types of tables are given in the AASTeX guidelines at \url{http://journals.aas.org/authors/aastex.html}

\section{Astrometric Selection of Probable Complex Members}\label{sec:selection}

The unprecedented astrometric precision of Gaia DR3 enables the identification of comoving groups and parallax measurements at the distance of G352 and beyond. This allows us to construct a high-confidence set of probable complex members consisting of young X-ray emitting stars, YSOs, and OB stars physically associated with the GMC. Our procedure involves first identifying clusters of sources with common proper motions, and then cleaning the sample of objects with discrepant parallaxes. While this Gaia-selected sample will be highly incomplete, because the MOXC and SPICY catalogs penetrate much more deeply into the obscured and embedded stellar populations within the GMC, the number of astrometrically-identified members is large enough to provide statistically robust information about the large-scale 3D structure and dynamics of the underlying G352 stellar population.

\subsection{Identification of Large-Scale Proper Motion Groups}

We performed all of our proper motion group identification in the ICRS frame, because it is the native reference frame for the proper motion uncertainties reported in the Gaia DR3 catalog. After identifying groups, we converted proper motions to Galactic coordinates for further analysis and interpretation.

We first divided the G352 initial sample into three large spatial regions according to Galactic longitude: NGC 6357 ($l\ge 352.6\degree$), IRDC bridge ($352.6\degree>l>351.8\degree$), and NGC 6334 ($l\le 351.8\degree)$.
For NGC 6357 and NGC 6334, we identified proper motion (PM) groups as follows:
\begin{enumerate}
    \item We computed the uncertainty on the two-dimensional proper motions for each Gaia DR3 source as $\sigma_\mu=\sqrt{\texttt{pmra\_error}^2+\texttt{pmdec\_error}^2}$. We then calculated the median uncertainty separately for sources in NGC 6357 (Med$[\sigma_\mu] = 0.63$ mas/yr) and NGC 6334 (Med$[\sigma_\mu] = 0.45$ mas/yr).
    \item We applied the cluster-finding algorithm DBSCAN \citep{DBSCAN96} to the distributions of \texttt{pmra} and \texttt{pmdec} in each spatial region. We identified PM groups of ${\ge}12$ sources, using $\sigma_\mu/2$ as the \texttt{EPS} (separation) parameter.
\end{enumerate}
The above procedure identified a single PM group in NGC 6334. In NGC 6357, three PM groups were identified: one large plus two smaller groups, the latter tightly clustered and well-separated from the main G352 population in PM space (Figure~\ref{fig:PMgal}). We refer to these smaller PM groups as Foreground 1 and 2. 
%{\it Given the complex spatial dispersion and extent, we will call these ``groups" in general, since they are not all spatial ``clusters".} 

Compared to NGC 6357 and NGC 6334, the IRDC bridge region contains an order of magnitude fewer Gaia-selected sources, and these are more highly dispersed in both space and proper motion. DBSCAN group-finding was not feasible to identify candidate members associated with the IRDC. Since the IRDC is located spatially between the two large complexes, and their PM distributions strongly overlap, we nominated sources for the IRDC PM group if both \texttt{pmra} and \texttt{pmdec} lay within 2 standard deviations of the NGC 6357 group locus or 3 standard deviations of the NGC 6334 locus (see Figure~\ref{fig:PMgal}).

\subsection{Parallax Zero-Point Corrections}

Parallax measurements in Gaia DR3 are subject to systematic uncertainties due to zero-point offsets. These offsets depend on the sky position, color, and magnitude of an individual source. Because the sources in our sample suffer heavy extinction and reddening, we found that fewer than half fell within the ranges of Gaia colors and magnitudes used by \citet{DR30pt} to calibrate  zero-point corrections ($N(Z_5 \cup Z_6)/N_{\rm tot}=0.46$). For sources that could be directly calibrated, we computed either the 5- or 6-parameter zero-point corrections and applied them to the DR3 parallax. For the remaining sources we applied the mean zero-point correction value ($\langle Z_5 \cup Z_6\rangle$) of all calibrated sources within the parent spatial region (NGC 6357, NGC 6334, or IRDC; Table~\ref{tab:0pt}). To account for the loss of precision in these uncalibrated zero-point corrections, we computed a systematic uncertainty using the as the standard deviation of observed corrections, $\sigma(Z_5\cup Z_6$), and added this (in quadrature) to the parallax uncertainties for the uncalibrated sources.
All subsequent analysis used these zero-point corrected parallaxes (denoted $\varpi^\star$) with uncertainties ($\sigma^\star_\varpi$).

\begin{deluxetable}{lcccc}\label{tab:0pt}
\centering 
\tablecaption{Parallax Zero-point Corrections by Location Within G352}
\tablehead{
\colhead{Region} &
\colhead{$N(Z_5 \cup Z_6)$} &
\colhead{$N_{\rm tot}$} &
\colhead{$\langle Z_5 \cup Z_6\rangle$ (mas)} &
\colhead{$\sigma(Z_5 \cup Z_6)$ (mas)}
}
\startdata
NGC 6357 & 560 & 1312 & -0.027 & 0.021 \\
NGC 6334 & 548 & 1087 & -0.028 & 0.020 \\
G352 IRDC & 65 & 120 & -0.026 & 0.020 \\  
\enddata
\tablecomments{See \citet{DR30pt} for definition and calculation of these zero-point corrections.}
\end{deluxetable}

\subsection{Parallax Cleaning to Produce Final Membership Sample}
As a final step to clean our membership sample, we identified and removed any sources that were included in one of our proper motion groups but had significantly discrepant parallaxes. First, we 
computed the median parallax ($\varpi^\star$, corrected for zero-point offsets) of each PM group. We then reassigned individual sources to the unclustered, ``field'' population if their parallaxes deviated by $>3\sigma^\star_\varpi$ from the median parallax. In all, we removed 37 sources from NGC 6357, 31 from NGC 6334, and 7 from the IRDC on the basis of discrepant parallaxes.
Because the two foreground groups superimposed on NGC 6357 are highly dispersed spatially, we took the additional step of removing any sources with \texttt{parallax\_over\_error}$< 3$ from these groups and reassigning them to the field population. 
%We did not find any sources with parallaxes consistent with membership that were not already included among the PM groups. 

\section{Results and Discussion}\label{sec:results}
The final, cleaned sample contains 1727 probable members of the G352 complex and 40 stars in the two foreground groups (Table~\ref{tab:results}). 
For each PM group (or spatial cluster, see Section~\ref{sec:substructure} below) we provide the number of stars, central Galactic coordinates, angular radius ($R_{3\sigma}$, enclosing 3 standard deviations from the central coordinates), mean Galactic PM $(\mu_l,\mu_b)$ and associated radius ($R_{3\sigma_\mu}$, enclosing 3 standard deviations from the mean PM), median corrected parallax with corresponding distance $d=1/{\rm Med}(\varpi*)$ and physical radius. 
Taking the weighted mean and standard deviation of median parallaxes across the three large-scale PM groups, we obtain an average heliocentric distance of $1670\pm 80$~pc for the G352 GMC complex. 

%Table of PM Groups
% Group     N  (l,b)  PM(Gal, deg/Myr) sigma_PM   Med parallax  Sigma parallax Distance (pc) 
%                                   
\begin{deluxetable}{rrcrccccc}
\label{tab:results}
\centering 
\tablecaption{Proper Motion Groups and Spatial clusters}
\tablehead{
\colhead{Group/} &
\colhead{$N$} &
\colhead{$(l,b)$} &
\colhead{$R_{3\sigma}$} &
\colhead{$(\mu_l,\mu_b)$} &
\colhead{$R_{3\sigma_{\mu}}$} &
\colhead{Med($\varpi*$)} &
\colhead{$d$} &
\colhead{$R_{3\sigma}$}
\\
\colhead{Cluster} &
\colhead{} &
\colhead{(\deg)} &
\colhead{(\arcmin)} &
\colhead{(mas yr$^{-1})$} &
\colhead{(mas yr$^{-1})$} &
\colhead{(mas)} &
\colhead{(pc)} &
\colhead{(pc)}
}
\startdata
   \multicolumn{9}{l}{NGC 6357:} \\
All        & 902  &  (353.15,+0.74)  &  22 & $(-2.39\pm 0.02,-0.56\pm 0.02)$  & 1.90  
    & $0.60\pm 0.03$  & $1670^{+90}_{-60}$ & 11 \\
Pismis 24      & 305  &  (353.15,+0.88)  &   5 & $(-2.16\pm 0.04,-0.49\pm 0.03)$  & 1.81
    & $0.59\pm 0.04$  & $1690^{+140}_{-110}$ & 2.5 \\   
G353.1+0.6 & 230  &  (353.09,+0.62)  &   5 & $(-2.33\pm 0.04,-0.57\pm 0.04)$  & 1.82
    & $0.58\pm 0.05$  & $1710^{+170}_{-140}$ & 2.5 \\
G353.2+0.6 & 145  &  (353.24,+0.63)  &   4 & $(-2.64\pm 0.06,-0.69\pm 0.04)$  & 1.83
    & $0.64\pm 0.07$  & $1570^{+200}_{-160}$ & 1.8 \\
\hline 
G352 IRDC  &  69  &  (352.23,+0.61)  &  42 & $(-1.85\pm 0.11,-0.91\pm 0.08)$  & 2.37
    & $0.60\pm 0.10$  & $1670^{+320}_{-230}$ & 20 \\
\hline 
   \multicolumn{9}{l}{NGC 6334:} \\
All        & 756  &  (351.09,+0.72)  &  49 & $(-1.62\pm 0.02,-1.03\pm 0.02)$  & 1.78
    & $0.59\pm 0.03$  & $1690^{+140}_{-110}$ & 24 \\
G351.2+0.7 & 219  &  (351.24,+0.67)  &  10 & $(-1.54\pm 0.04,-1.13\pm 0.04)$  & 1.63
    & $0.60\pm 0.05$  & $1680^{+140}_{-130}$ & 4.9 \\
G351.0+0.7 &  87  &  (351.04,+0.66)  &   5 & $(-1.42\pm 0.05,-0.90\pm 0.06)$  & 1.51
    & $0.57\pm 0.06$  & $1760^{+210}_{-170}$ & 2.5 \\
GM 1--24    &  58  &  (350.51,+0.95)  &   5 & $(-1.77\pm 0.07,-1.32\pm 0.07)$  & 1.66
    & $0.52\pm 0.09$  & $1940^{+410}_{-310}$ & 2.8 \\
\hline
Foreground 1       & 30   &  (353.12,+0.62) &  35 & $(-0.91\pm 0.05,-2.87\pm 0.04)$  & 0.79
    & $0.99\pm 0.04$  & $1000^{+50}_{-40}$ & 10 \\
Foreground 2       & 10   &  (353.04,+1.21) &  8  & $(-1.49\pm 0.04,-4.03\pm 0.04)$  & 0.42
    & $0.99\pm 0.06$  & $1010^{+70}_{-60}$ & 2.4 \\
\enddata
\end{deluxetable}
\begin{figure}[ht!]
\plotone{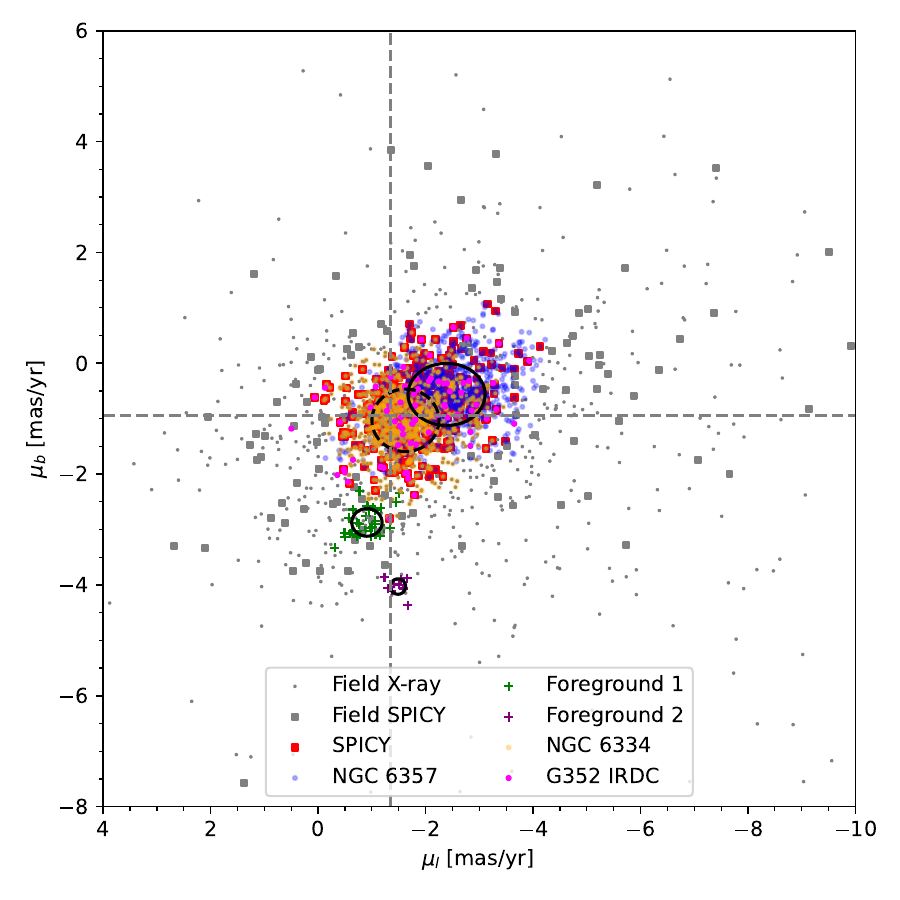}
\caption{Proper motion plot in Galactic coordinates showing sample of probable members (symbols colored according to spatial/proper motion groups defined in the text) and field stars (gray symbols). The dashed vertical and horizontal lines intersect at the expected proper motion of sources at the G352 distance, assuming Galactocentric circular rotation (see text). Solid ellipses enclose the 1-sigma locus for sources associated with the three proper motion groups in NGC 6357; dashed ellipse shows the same for all sources associated with NGC 6334. 
\label{fig:PMgal}}
\end{figure}

The results of our PM membership selection are illustrated in Figure~\ref{fig:PMgal}, in which Gaia DR3 ICRS proper motions have been converted to Galactic coordinates using \texttt{astropy.coordinates}. The large PM groups associated with NGC 6357 (blue) and NGS 6334 (orange) overlap substantially; while it might be possible to statistically identify separate peaks in PM space, the assignment of many individual members of each group cannot be done based on PM alone. By construction, the G352 IRDC population (pink) spans the full range of PM space occupied by NGC 6357 and NGC 6334. The two foreground groups are well-separated from G352 in PM space.

%\todo{SKIP FOR NOW: Introduce choice of PM reference frame, co-rotating with circular orbit at Galactocentric radius of G352, and how we corrected for Sun's peculiar motions [CITATIONS???! Why did I misplace my notes??] converted to this frame (Figures \ref{fig:full_spatPM} and \ref{fig:subclusters}). }
To define a physically-motivated reference frame for our subsequent analysis and discussion of the kinematics, we computed the proper motion of a circular orbit at the Galactocentric location of the G352 complex, based on our revised average heliocentric distance to G352 and $R_0=8180$~pc \citep{GRAVITY19}. We used the Galactic rotation curve of \citet{reid19}, which includes the Sun's peculiar motion of $(U_\odot,V_\odot,W_\odot)=(10.6,10.7,7.6)$~\kms\ with respect to a local standard of rest orbital velocity of $\Theta_0=236$~\kms. This predicts a circular orbit proper motion vector with $(\mu_{l,c},\mu_{b,c})=(-1.34,-0.94)$~mas~\peryr\ (dashed lines in Figure~\ref{fig:PMgal}).

%of $(u_\odot,v_\odot,w_\odot)=(11.1,12.24,7.25)$~\kms\ \citep{SBD10} to compute the proper motion of a circular orbit at the Galactocentric location of the G352 complex. We used the Galactic rotation curve of \citet{eilers19} with $V_0=229$~\kms, $R_0=8180$~pc \citep{GRAVITY19}, and our revised heliocentric distance to G352. This yielded circular orbit proper motions of $(\mu_{l,c},\mu_{b,c})=(-1.31,-0.90)$~mas/yr (dashed lines in Figure~\ref{fig:PMgal}).

%\tbr{MIKE: Given that you used the V0 value of from Eilers+2022, I was wondering whether any of the scientific results would change in a notable way if you instead used the slightly faster V0=236 km/s from Reid+2019, which was fitted to younger tracers of Galactic rotation? SIMRAN is recalculating this for us... $\mu_l = -1.34,\mu_b=-0.94$ relative to Galactic Center. \citet{reid19} Column A5 values...for rotation curve and peculiar motions.}

\subsection{Large-Scale Structure and Kinematics of the G352 Complex}

\begin{figure}[ht!]
\hspace{0.36in}\plotone{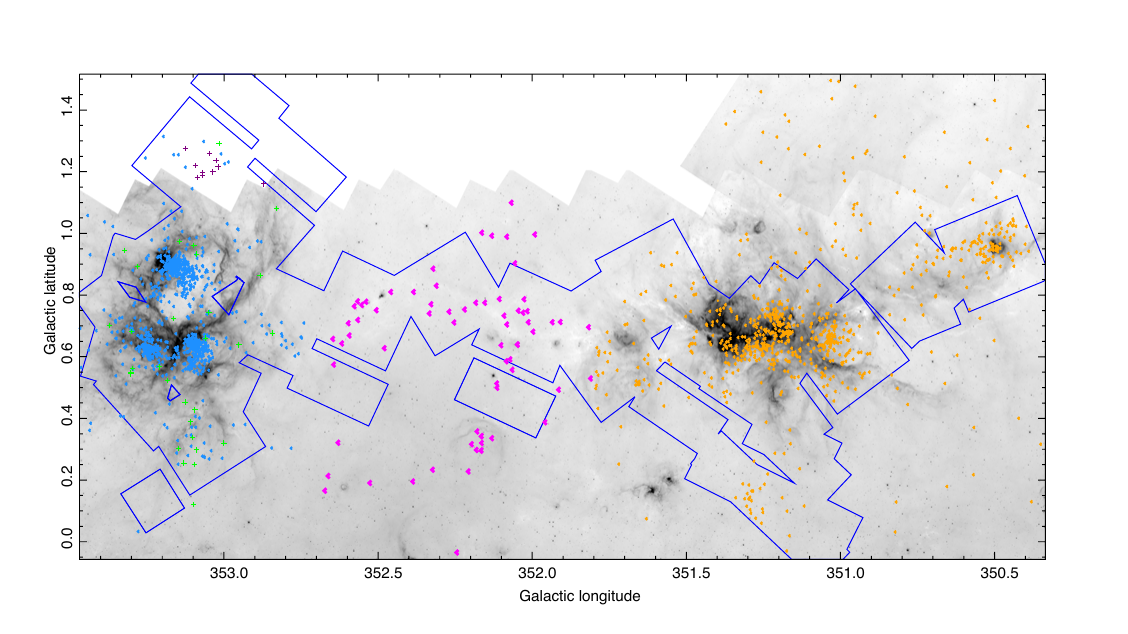}
\plotone{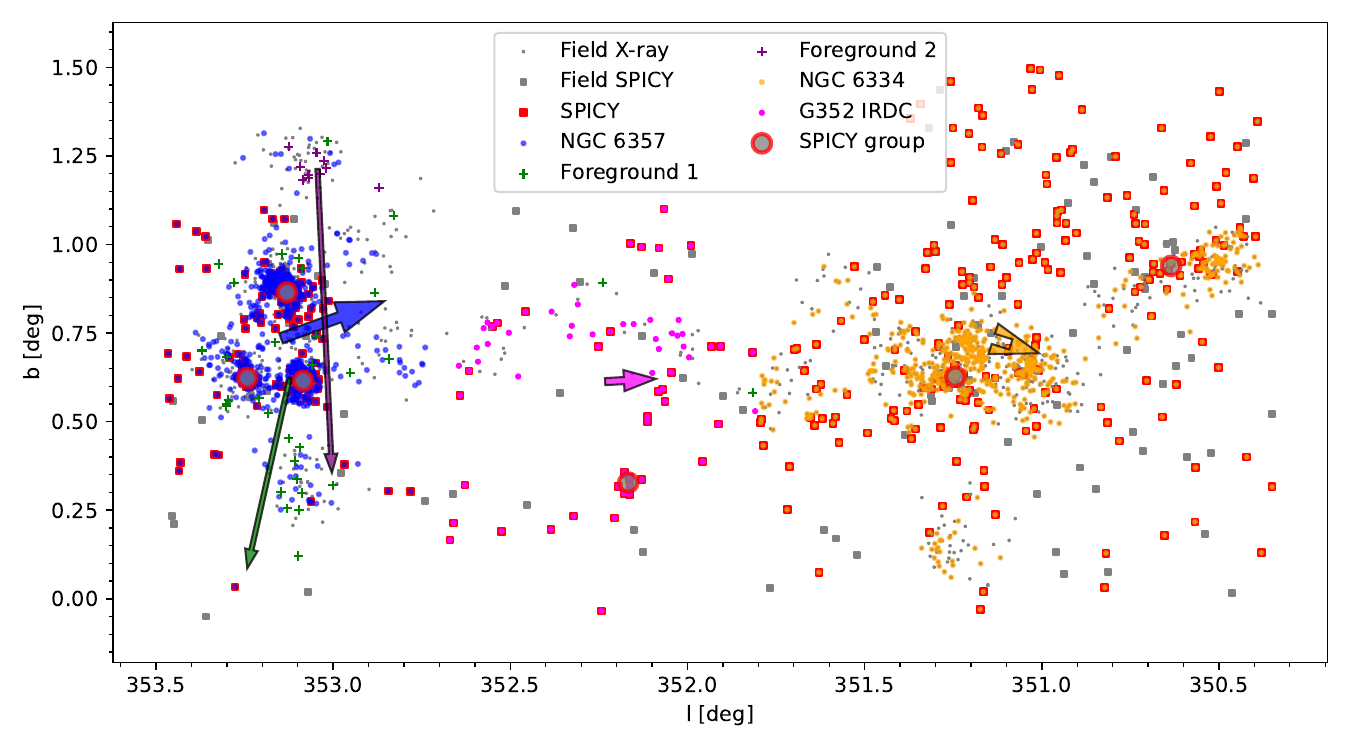}
\caption{{\it Top:} Membership sample, color-coded by PM group (as in Fig.~\ref{fig:PMgal}), overlaid on a GLIMPSE 8.0~\um\ mosaic (inverted logarithmic grayscale) of the G352 complex. The blue polygons outline the footprint of the ACIS-I mosaic.
{\it Bottom:}  Membership sample plotted in Galactic coordinates, plus field sources removed from the sample on the basis of discrepant PM or parallax (gray symbols). Arrows show the mean proper motion of each PM group in Table~\ref{tab:results} over the 1~Myr typical age of the stellar clusters, relative to a Galactic circular orbit.
\label{fig:full_spatPM}}
\end{figure}

Two visualizations show the spatial distribution of the 1767 sources in our final, cleaned sample
(Figure~\ref{fig:full_spatPM}). In the top panel, we overlay the source positions on a Spitzer/GLIMPSE 8.0~\um\ mosaic \citep{benjamin_2003}, to illustrate the relationships between the various clusters and groups with dusty ISM structures and the Spitzer and Chandra survey footprints. In the lower panel, we plot the sources with proper motion vectors representing the movement of each large-scale PM group, relative to the Galactic circular orbit, projected 1~Myr forward in time \citep[comparable to the current age of the G352 stellar populations;][]{AgeJX}.
While this sample is statistically large enough to reveal the overall kinematics of the G352 constituent populations, it  represents a small fraction of the true stellar population. Our Gaia selection criterion, requiring detection at visible wavelengths, is the most stringent limitation given the high level of dust extinction produced by the GMC itself. It is likely that the majority of the X-ray point sources and SPICY objects found within our study area that do not have Gaia counterparts are also members of the G352 complex. A study of this larger population is beyond the scope of this work.

The three massive clusters of NGC 6357 are readily apparent in Figure~\ref{fig:full_spatPM}, and all nestle within 8~\um\ nebular shell structures delineating photodissociation regions. 
The G352 IRDC bridge filament is traced by X-ray-identified stars along its entire length. The region where the IRDC meets NGC 6334 is known to be a very active site of ongoing star formation \citep{russeil10,MIRES,willis13} but here the incompleteness of our Gaia-selected sample may be maximal due to extremely high obscuration in the dense molecular cloud.
Stars associated with the bright 8~\um\ filament of NGC 6334 present a chaotic spatial structure, with many dense concentrations and also a distributed population. GM1-24  \citep{GM77} is spatially distinct from the main body of the NGC 6334 PM group. It hosts a cluster of X-ray members within its compact \hii~region, with an adjacent SPICY cluster in a dark region of the filament.

The Gaia proper motions (Figures~\ref{fig:PMgal} and \ref{fig:full_spatPM}) reveal that the stellar populations in the G352 complex orbit the Galactic center more slowly than predicted by the circular rotation model.
%NGC 6334 and the IRDC bridge generally follow Galactic circular motion. 
The deviation is most pronounced for NGC 6357, with $\mu_l -\mu_{l,c}=-1.05$~mas~\peryr. The corresponding peculiar velocity component lags the circular orbital speed by $-8.36$~\kms\ (parallel to the plane). NGC 6334, by contrast, has a much smaller peculiar velocity. Comparing NGC 6357 to NGC 6334, $\Delta\mu_l=-0.77$~mas~\peryr, or $-6.13$~\kms. Perpendicular to the plane, NGC 6357 is moving away from the midplane (toward Galactic north) while NGC 6334 is moving in the other direction, with $\Delta{\mu_b}=0.47$~mas~\peryr, for a peculiar velocity amplitude of 3.73~\kms. Another indication that the entire G352 filament lags the circular orbit model is  present in the CO radial velocity \citep{fukui18a}. The circular orbit model predicts $v_{r,c}=-6.5$~\kms, but the bulk of the CO is observed in the range $v_r = 2$--4 \kms, with even larger deviations observed where the IRDC filament meets NGC 6357.

%\tbr{MIKE: it might help to be careful with what you mean by undulation and deceleration. For deceleration do you mean some dv/dt, or do you mean a spatial velocity gradient?}

\subsection{The Foreground Groups: Interlopers from SCO OB4?}
The median parallaxes of the two foreground PM groups  place them at heliocentric $d=1$~kpc (Table~\ref{tab:results}). While their median proper motions differ from each other by more than $3\sigma$,  both are falling toward the Galactic midplane; at $-6.8$~\kms\ and $-12.1$~\kms\ for Foreground 1 and 2, respectively. Foreground 1 is spatially dispersed across the entire NGC 6357 area on the sky, while Foreground 2 clusters within the single ACIS-I pointing in our mosaics that happened to fall just above the latitude coverage of the GLIMPSE (and hence SPICY) survey area (Figure~\ref{fig:full_spatPM}). Both foreground clusters were identified on the basis of X-ray, not YSO, detections. The lack of YSO members (no evident infrared excess from dust disks) suggests that Foreground 1 consists of diskless pre-main-sequence stars that are older than the G352 population. We cannot say the same for Foreground 2, as we have no information from Spitzer about a presence or absence of YSOs in this group.

The foreground groups could be associated with the widely-dispersed and often-overlooked Scorpius OB4 association, which lies ${\sim}2\degree$--$4\degree$ degrees above the Galactic midplane near G352 \citep{AG76}. Using Gaia DR2 astrometry, \citet{GDR2_OB} identified 11 SCO OB4 members scattered across $351.5\degree < l < 353.5\degree$ and $2.5\degree<b<4.5\degree$;
with mean proper motion $(\mu_l,\mu_b)=(-1.34\pm 0.42,-2.70\pm 0.29,)$~mas~yr$^{-1}$ and $d=0.96$~kpc. Foregrounds 1 and 2 occupy a similar region of PM space at the same distance (Table~\ref{tab:results}); if they are part of SCO OB4 then this association extends toward lower Galactic latitudes than previously mapped, passing in front of NGC 6357 but not the IRDC or NGC 6334. 

%\todo{Paragraph noting the very different PM and spatial distirbution of the two FG groups. Can also mention the spatially distributed population, and several far-flung spatial clusters that do have PM and parallax consistent with membership, but association is questionable/surprising given large perpendiculat distance from main filament.}

A spatially distributed population of SPICY YSOs extends high above NGC 6334 (Figure~\ref{fig:full_spatPM}). The proper motions and parallaxes of these sources are, like NGC 6334, similar to Galactic circular orbits at the G352 distance and hence inconsistent with SCO OB4. Below the IRDC filament, a SPICY group at $(l,b)=(352.15\degree,0.35\degree)$ is surrounded by an 8~\um\ arc structure indicating a young cluster powering an \hii region bounded by a photo-dissociation region. While far removed spatially from the G352 filament, the parallax and proper motion of this SPICY group are indistinguishable from the G352 IRDC population.
%The other is a group of X-ray sources at $(l,b)=(351.3\degree,0.15\degree)$, ...

\subsection{Inter-Cluster Kinematics and 3-D Structure within NGC 6357 and NGC 6334}\label{sec:substructure}

%\subsection{Spatial Clusters within Large-Scale Proper Motion Groups}

%\todo{Both NGC 6357 and NGC 6334 exhibit complex spatial substructures...Pismis 24 and three other major cluster [CITATIONS] and NGC 6334 is much more complex, and GM1-24 is really a separate cloud [CITATIONS]. Cannot distinguish using PM alone, so...}

While the stellar populations within both NGC 6357 and NGC 6334 exhibit complex spatial structure, it was not possible to identify sub-groups on the basis of proper motion alone. We again applied DBSCAN, this time to identify spatial clusters within each large PM group. The separation parameter \texttt{(EPS)} was defined as the typical separation on the sky for stars within each PM group: $\epsilon = A/\sqrt{N}$, with $A$ approximating the area on the sky (in square degrees) occupied by each large PM group and $N$ defining the minimum cluster size to identify. 
We used $A=0.75$~deg$^2$, $N\ge 15$ for NGC 6357 and $A=1$, $N\ge 20$ for NGC 6334. 
The large majority (75\%) of NGC 6357 members reside in one of the three rich spatial clusters, Pi 24, G353.1+06, or G353.2+0.6 (Table~\ref{tab:results}). In contrast, slightly over half of stars associated with NGC 6334 (52\%) were not assigned to any of its three identified large spatial clusters.
\begin{figure}[ht!]
\includegraphics[height=2.6in]{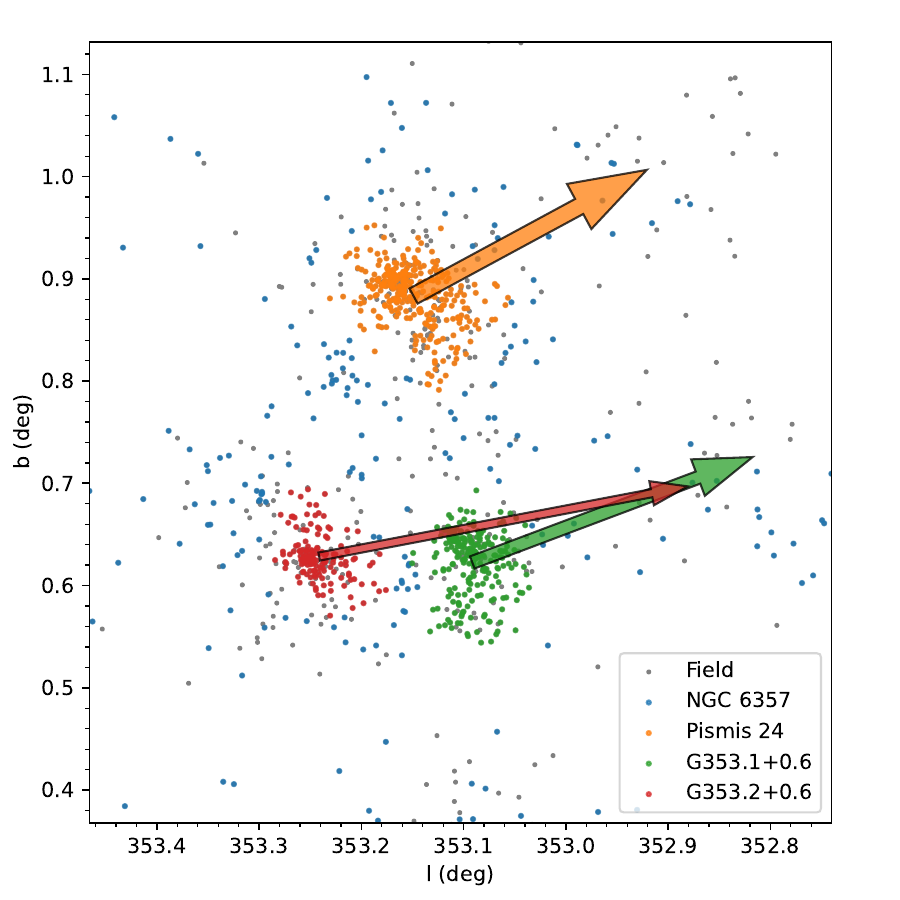}
\includegraphics[height=2.6in]{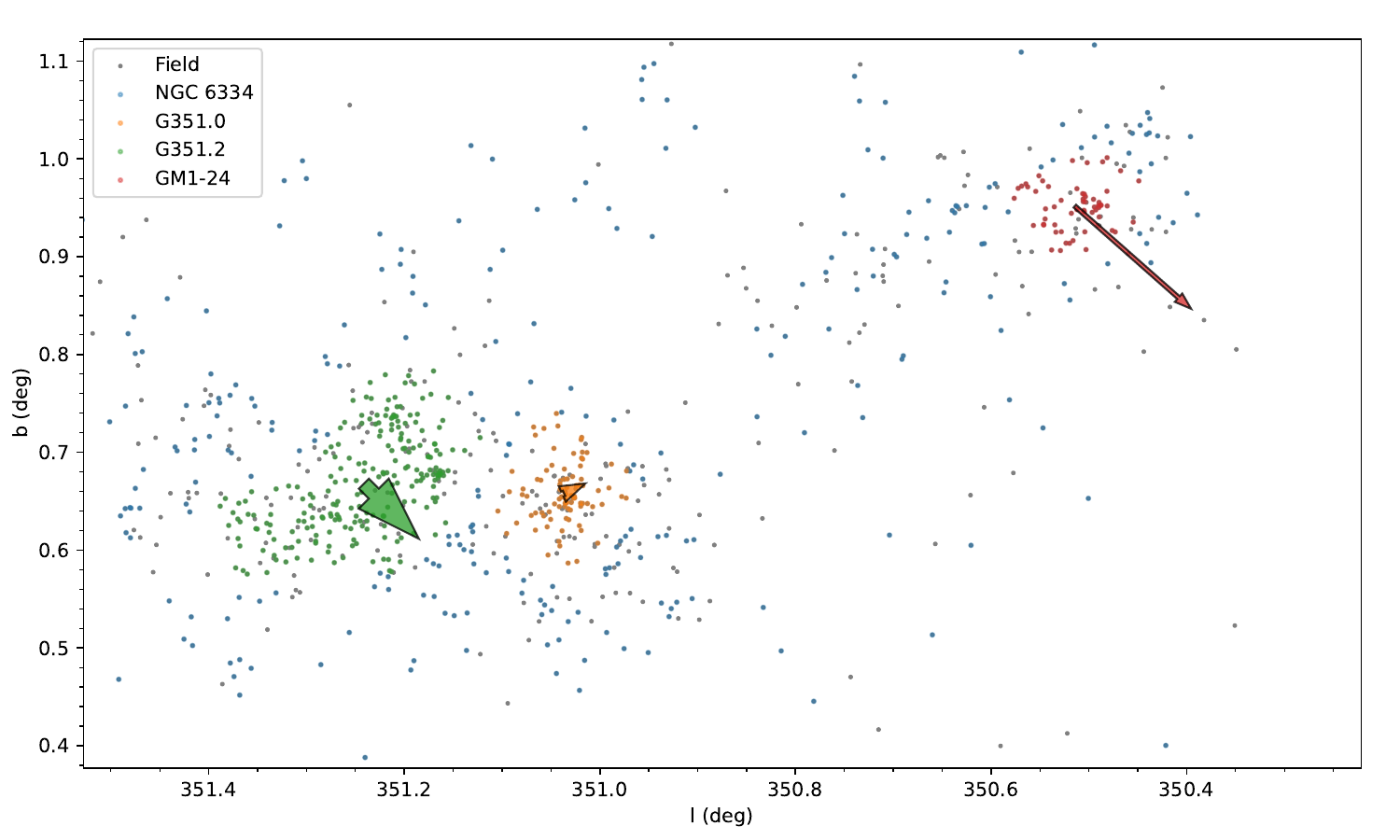}
\caption{Relative motions of spatial clusters within each complex ({\it Left:} NGC 6357 and
{\it Right:} NGC 6334). Arrow lengths show proper motions over 1 Myr relative to Galactic rotation, while the arrow widths are scaled to the cluster sizes (as in Figure~\ref{fig:full_spatPM}).
\label{fig:subclusters}}
\end{figure}
The membership and relative proper motions of the three spatial clusters in each of NGC 6357 and NGC 6334 are plotted in Figure~\ref{fig:subclusters} and  included in Table~\ref{tab:results}. These results agree qualitatively with the picture presented by \citet[][their Fig.\ 14]{russeil20} using Gaia DR2 astrometry for YSO groups.

In NGC 6357, we find that all three clusters contribute to the overall peculiar motion with respect to a Galactic circular orbit; they are moving too slowly in the orbital direction while rising away from the midplane. Pi 24 and G353.1 have statistically indistinguishable parallaxes and nearly parallel proper motions, they appear to be neighboring clusters that are gradually diverging. By contrast, G353.2, which appears closer to G353.1 on the sky, has a larger median parallax that deviates from the other two clusters at the $1\sigma$ level, as well as a larger proper motion. This cluster may be in front of the rest of NGC 6357, placing it at the leading end of the GMC filament in Galactic orbit. Even if all three clusters are at the same distance, their proper motions all diverge, so they are unlikely to be gravitationally bound to each other.

The two spatial clusters in NGC 6334 are G351.2 and G351.0, with median parallaxes consistent with the average GMC distance, given the uncertainties. Their motions deviate significantly from Galactic circular orbits, but the peculiar velocities are much less than those of the three NGC 6357 clusters. The smaller of the two clusters, G351.0, shows a dense core in the Gaia-selected sample thanks to its relatively low obscuration within a feedback-evacuated, broken bubble structure on the trailing edge (in Galactic rotation) of the GMC filament. The directions of the PM vectors for each of these clusters diverge, hence they are not likely gravitationally bound to one another.

GM 1--24 is the third spatial cluster identified within the NGC 6334 PM group. The nebular morphology of G352 in cold dust and PAHs (see Figure~\ref{g352fullfield.fig} as well as CO velocities \citep{fukui18a}
strongly suggests a continuous filament connecting GM 1--24 to the main body of NGC 6334. The median parallax of GM 1--24 places it at a greater distance than the main complex, although the uncertainties are relatively large. Its proper motion deviates significantly from the other two NGC 6334 clusters (Figure~\ref{fig:PMgal}).

\subsection{G352 in the Galactic Context}
\begin{figure}[ht!]
\centering
\includegraphics[width=6in]{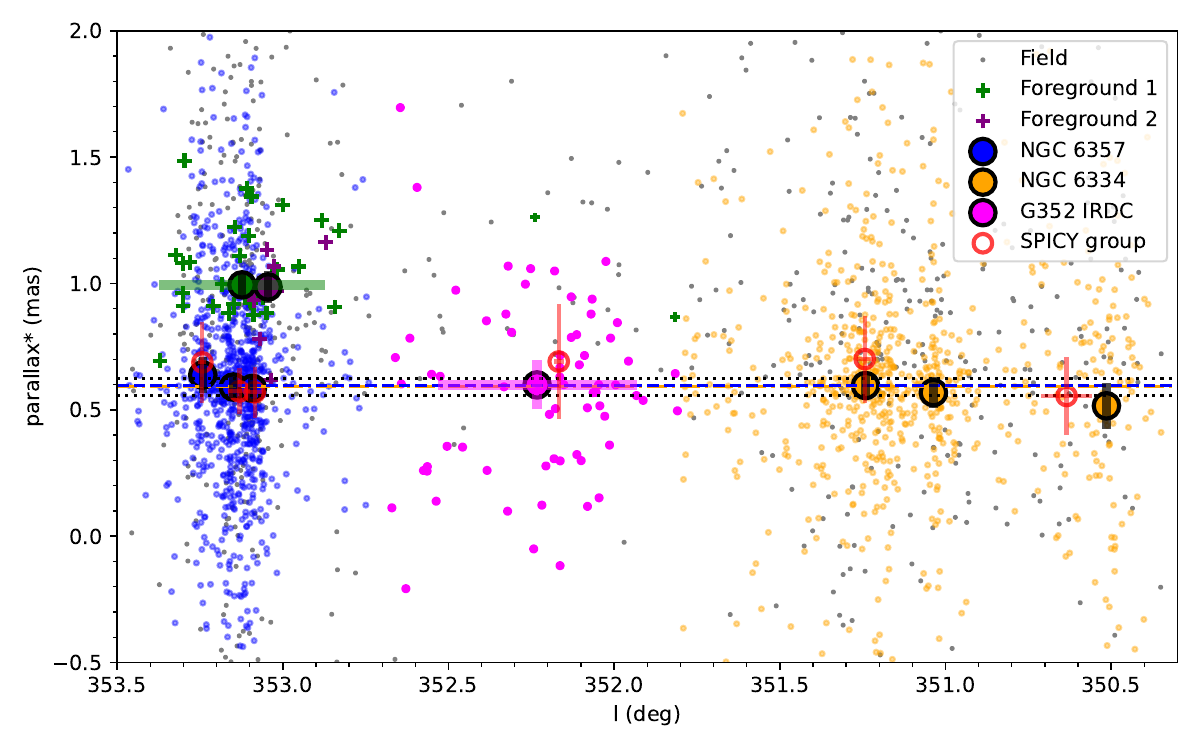} \\
\includegraphics[width=5in]{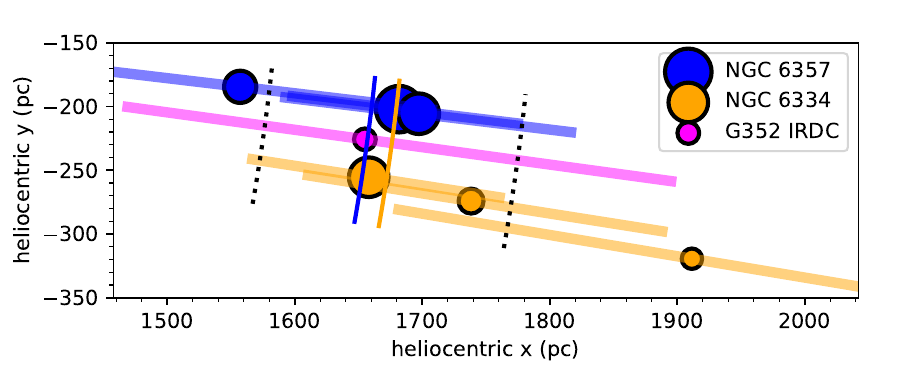}
\caption{{\it Top:} Parallax versus Galactic longitude for all sources in our initial sample. Symbols and colors are the same as in Figure~\ref{fig:full_spatPM}. {\it Bottom:} Group parallax distance versus physical separation in Galactic $(x,y)$ coordinates. The two foreground groups are cropped out of this view.
\label{fig:distances}}
\end{figure}

The corrected parallaxes of every Gaia DR3 source used in this study are plotted against Galactic longitude in the top panel of Figure~\ref{fig:distances}. The median parallaxes and uncertainties for each PM group and spatial cluster from Table~3 are overplotted and color-coded to match their constituent stars. The individual parallaxes scatter widely around the median of each group, but thanks to the statistically large samples used, we are able to determine the median parallaxes with unprecedented accuracy. This begins to reveal the 3-dimensional structure of G352, hinting at a depth into the Galactic plane comparably expansive as its panoramic, 150-pc sweep across the sky (Figure~\ref{g352fullfield.fig}).
In the bottom panel of Figure~\ref{fig:distances} we provide a bird's eye view of the layout of G352 in physical coordinates, as it might appear to an extragalactic observer looking down from Galactic north. While the distance uncertainties of adjacent  spatial clusters overlap, there is a general trend of increasing distance ($x$) with decreasing Galactic longitude ($y$). These results, based on associated stellar populations, suggest that the G352 filament itself is tilted into the sky with a significant pitch angle. It may also contain bends or kinks, or be broken up by feedback-driven bubbles between overlapping segments in some places, for example in NGC 6357/Pi 24, where the effects of radiation pressure and winds from massive stars are strongest.

The high-resolution DECaPS 3D dust maps presented by \citet{Z25} provide independent evidence that G352 is part of a high pitch-angle filament. 
%These dust maps, constructed using a wholly independent methodology leveraging foreground and background stars, appear fully consistent with our investigation of the stellar populations associated with G352. 
The DECaPS maps show a tilted dust filament spanning a much larger range of Galactic longitude ($345\degree<l<360\degree$) than covered by our contiguous X-ray survey, including the RCW 120 and NGC 6231 \hii regions further south.  G352 has historically been considered part of the Sagittarius-Carina spiral arm, and while we do find our average distance and longitude coincide with the location of the predicted arm \citep{reid19}, the high pitch angle of the G352 filament cannot fit with a lognormal density wave model. 
%We hence agree with \citet{Z25} that the spiral density wave models for the fourth Galactic quadrant likely need revising. 
G352 has a pitch angle opposite in orientation that of the Sagittarius Spur in the first Galactic quadrant \citep{Sagspur21}, such that the side leading the Galactic orbit is farther from the Galactic center. This suggests that the structure was not shaped by Galactic shear, which would stretch the GMC in the other direction.

%\todo{WEAK conclusion, instead take some of Mike's suggested discussion: ``I think it's notable that the structure is oriented in a way so that the outer side (relative to the Galactic Centre) is in front (relative to the direction of Galactic orbit). This pitch angle is opposite in orientation compared to the Sagittarius Spur, so I think it's worth pointing out. This suggests that this structure was not shaped by Galactic shear, which would stretch it out in the other direction."}

The kinematics of G352 may bear the imprint of dynamical processes that set the stage for the observed starburst activity. Essentially the entire length of the filament is  actively forming stars currently, but the largest burst commenced in NGC 6357 about 1.5~Myr ago  \citep{russeil10,AgeJX}, producing the the three most massive clusters. These three clusters exhibit the largest lags behind Galactic orbital speeds observed among the G352 population (Figure~\ref{fig:subclusters}). This raises the possibility that some event in the recent past, such as a cloud-cloud collision \citep[e.g.][]{fukui18a} or interaction with an expanding superbubble shell, impacted the leading end of the filament, compressing the gas and enhancing the local star formation rate. 
%\todo{Spiral density wave seems unlikely, superbubble shell?}

 Many previous studies \citep[e.g.][]{russeil10,fukui18a} have noted the apparent age gradient in star formation along the length of the G352 filament, from NGC 6357 (older) to NGC 6334 (younger), but the observed patterns of YSOs and X-ray pre-main-sequence stars paint a more complicated picture. For example, the G351.0+07 cluster in NGC 6334 may be older than Pi 24 \citep{AgeJX}.  Much of the IRDC bridge appears quiescent, but it shows signs of incipient cluster formation, including bright IR sources and YSOs.
While our parallax measurements of stars in the IRDC bridge are imprecise, this dense filament likely lies in front of the giant \hii regions, both of which produce diffuse X-ray emission that is clearly blocked by the IRDC, which is by definition an absorption feature in the mid-IR as well (see Figures~\ref{g352fullfield.fig} and \ref{g352.fig}). The CO velocity maps \citep{fukui18a} show that the IRDC is less blueshifted by ${\sim}2$~\kms\ compared to the bulk of the GMC, while stellar proper motions indicate that it is moving toward NGC 6334 at a similar relative velocity. Taken together, these observations suggest that the IRDC is converging with the leading edge of NGC 6334.  Where the IRDC connects to NGC 6334 the largest concentration of YSOs has been observed \citep{willis13,MIRES,SPICY}, revealing a very recent and ongoing burst of star formation.

\section{Conclusions}\label{sec:conclusions}
We have combined a large catalog of over 11,000 X-ray point sources, published here in its entirety for the first time, with the SPICY catalog to identify candidate young stellar members across the full $3\degree$ length of the G352 GMC complex. We focused this study on a subset of ${\sim}2000$ of these candidate young stars and YSOs that have visible-light counterparts with high-quality astrometry in the Gaia DR3 catalog. We used proper motions and spatial distributions to assign stellar members to the NGC 6357, NGC 6334, GM1-24 \hii regions, and to the IRDC bridge region. In the process, we also identified two foreground groups that may be part of the SCO OB4 association.

Our heliocentric distance to G352 of $d=1670\pm 80$~pc, using ${\sim}1700$ Gaia DR3 parallaxes to young stellar members, falls within the range of previous distance estimates from both gas kinematics \citep{russeil10, russeil16} and  Gaia DR2 parallaxes of the clustered OB and young stellar populations \citep{kuhn19,russeil20}. Our new distance is ${\sim}100$ pc less than the previous Gaia DR2 distances, the difference driven primarily by the systematic parallax zero-point correction implemented for DR3 \citep{DR30pt}.

Proper motions reveal significant velocity excursions among the G352 stellar clusters and groups. At the leading end of the filament (in the direction of Galactic rotation), three massive clusters in NGC 6357 all exhibit peculiar velocities that lag a Galactic circular orbit model by ${\sim}8$~\kms, while the peculiar velocities in NGC 6334 are much smaller. This deviation in kinematics may offer a clue as to why the morphologies of the two giant \hii~regions appear very different, in spite of forming from the same GMC complex at nearly the same time. NGC 6357 formed three monolithic, very massive clusters, the most famous being Pi 24. The stellar structure in NGC 6334 is far less organized, with few evident large clusters but numerous smaller ones and a larger fraction of spatially distributed stars. NGC 6334 retains a striking filamentary morphology, while any pre-existing filamentary structure in NGC 6357 is not apparent (Figure \ref{g352fullfield.fig}). We speculate that interaction with an expanding supperbubble shell or a similar collision with another cloud or shock front occurred 1--2~Myr ago and enhanced star formation on the leading edge of G352.
%The GMC filament appears to be tilted or curved into the sky, with stellar populations undu
%Investigations such as this one into young stellar kinematics over spatial scales of tens to hundreds of parsecs may offer new insights into the initial conditions distinguishing various modes of massive star formation activity observed in various GMC complexes. 

None of the various stellar proper motion groups identified with G352 are on paths that will converge with the others. While determining whether individual clusters such as Pi 24 will remain bound over long timescales is beyond the scope of this studey, we can conclude that the complex is not gravitationally bound at parsec scales or larger. Similar patterns of inter-cluster motions have been observed previously both in NGC 6357 and 6334 and in several other star-forming complexes \citep[e.g.][]{kuhn19,russeil20,kuhn26}. Looking ${\sim}10$ Myr into the future, G352 will become a large OB association, perhaps containing two or three massive open clusters and a similar number of co-moving groups, each setting out on its own path through the Galaxy.

\begin{acknowledgments}
%We appreciate the time and effort donated by our anonymous referee to comment on this paper.  
This work was supported by the {\em Chandra X-ray Observatory} Award
GO7-18003A   % AO18 (co-I's Broos, Povich)
(PI:  L. Townsley)
and by the Penn State ACIS Instrument Team Contract SV4-74018.  Both of these were issued by the \Chandra\ X-ray Center, which is operated by the Smithsonian Astrophysical Observatory for and on behalf of NASA under contract NAS8-03060.
The ACIS Guaranteed Time Observations included here were selected by the ACIS Instrument Principal Investigator, Gordon P. Garmire, of the Huntingdon Institute for X-ray Astronomy, LLC, which is under contract to the Smithsonian Astrophysical Observatory; Contract SV2-82024.
M.S.P. was also supported by the National Science Foundation through grant CAREER-1454333.
This research used data products from the \Chandra\ Data Archive, software provided by the \Chandra\ X-ray Center in the application package {\it CIAO}, and {\it SAOImage DS9} software developed by the Smithsonian Astrophysical Observatory. This work also used data from the ESA mission {\it Gaia} (\url{https://www.cosmos.esa.int/gaia}), processed by the {\it Gaia} Data Processing and Analysis Consortium (DPAC,
\url{https://www.cosmos.esa.int/web/gaia/dpac/consortium}); funding for the DPAC has been provided by national institutions, in particular the institutions participating in the {\it Gaia} Multilateral Agreement.   This research also used data products from the {\em Spitzer Space Telescope}, operated by the Jet Propulsion Laboratory (California Institute of Technology) under a contract with NASA.
%, data products from the {\em Wide-field Infrared Survey Explorer} ({\em WISE}), which is a joint project of the University of California, Los Angeles and JPL/CalTech, funded by NASA, and data products from the European Space Agency (ESA) {\em Herschel Space Observatory}, with science instruments provided by European-led Principal Investigator consortia and with important participation from NASA.  
 This research used NASA's Astrophysics Data System Bibliographic Services, and the SIMBAD database and VizieR catalog access tool provided by CDS, Strasbourg, France.

\end{acknowledgments}


\vspace{5mm}
\facilities{CXO(ACIS), Spitzer(IRAC), Gaia, CTIO:2MASS, ESO:VISTA}


\software{ACIS Extract \citep{broos_2010,AE12,AE22},
    CIAO \citep{CIAO},
	SAOImage ds9 \citep{ds9},
	TOPCAT \citep{TOPCAT},
    astropy \citep{2013A&A...558A..33A,2018AJ....156..123A},
    scikit-learn \citep{scikit-learn}
          }



\clearpage 

\appendix

\section{X-ray Sources in the G352 IRDC Bridge Observations}\label{sec:newdata}

In this Appendix we provide additional details of the Chandra/ACIS observations covering the G352 IRDC bridge between NGC 6357 and NGC 6334.
   
\begin{figure}[htb]
\centering
\plottwo{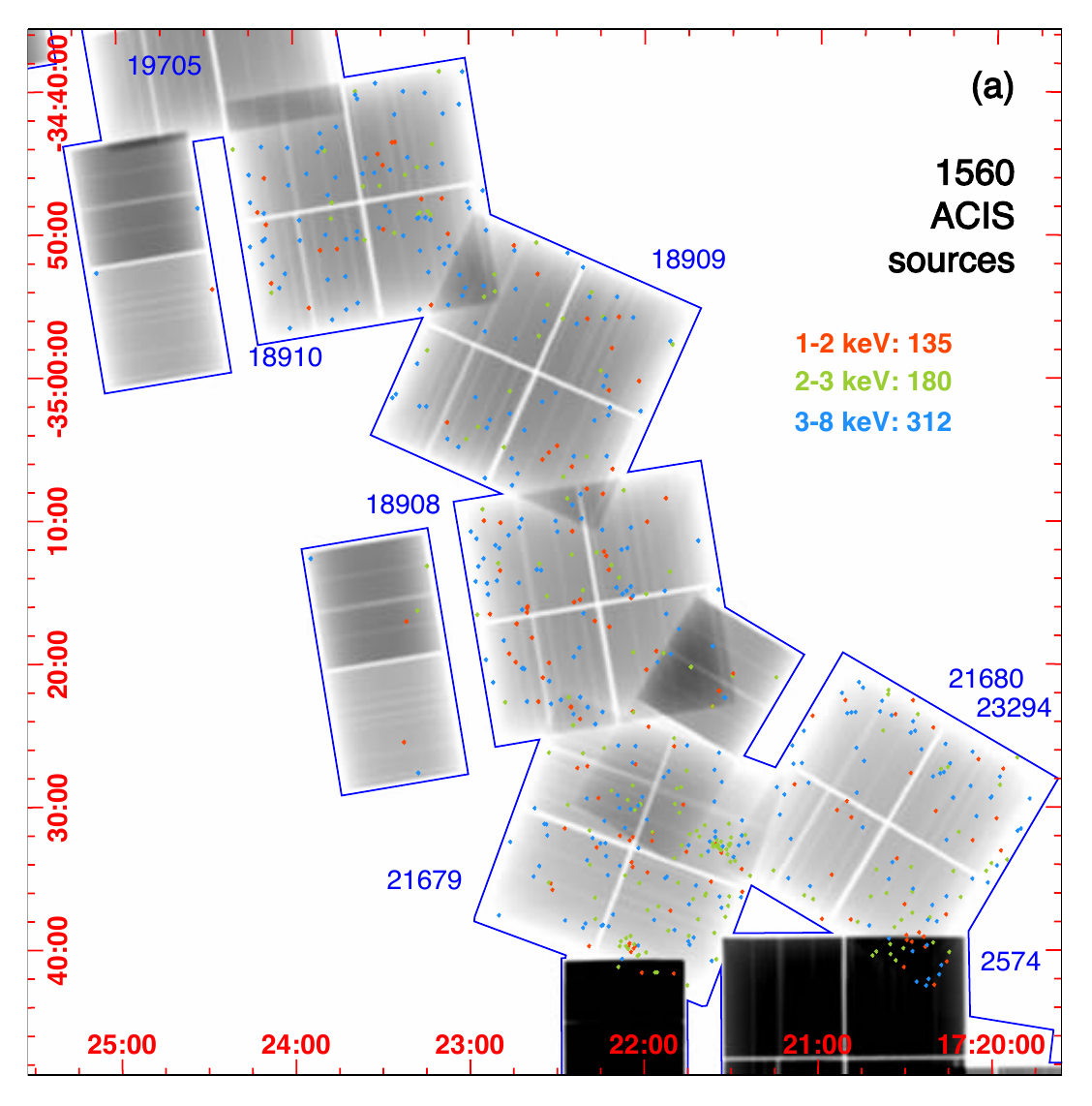}
{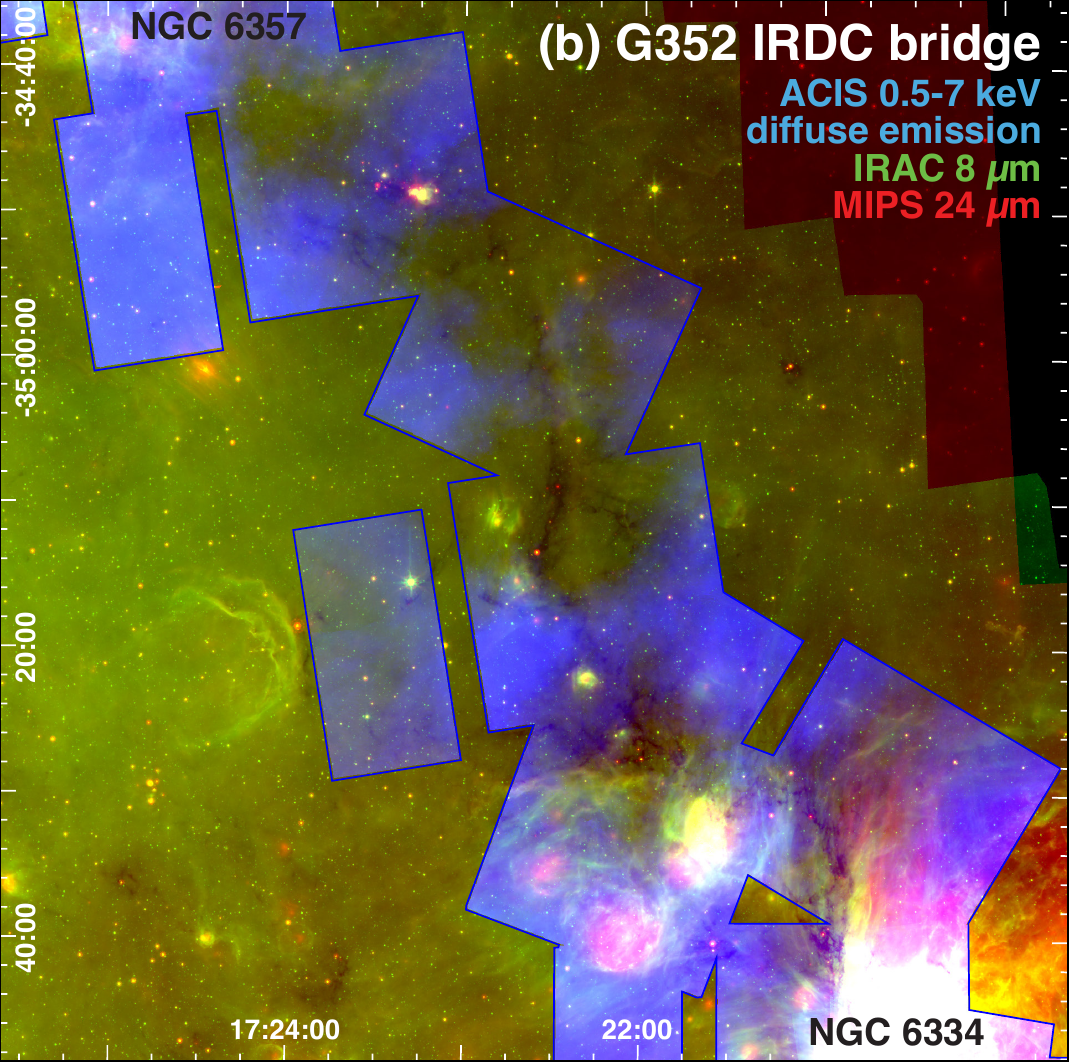}
\caption{Images presenting the new X-ray data on the G352 IRDC bridge between NGC~6357 and NGC~6334, displayed in equatorial coordinates.
(a) ACIS exposure maps for the G352 IRDC and bridge to NGC~6334 with brighter ($\geq$5 net counts) ACIS point sources overlaid; colors denote median energy for each source.  ObsID numbers are shown in blue (see Table~\ref{tbl:obslog}).  The total number of X-ray sources validated in our analysis is noted; the majority of faint sources are not shown here.
(b) X-ray diffuse emission combined with Spitzer MIR.
\label{g352.fig}}
\end{figure}

The observing log for the six Chandra observations in the G352 bridge is provided in Table~\ref{tbl:obslog}. The first three ObsIDs were obtained as part of a large GO program targeting four IRDCs, while the last three were obtained using Cycle 20 GTO time. All observations were in TE-VFAINT mode. Two visualizations of the G352 bridge observations are presented in Figure~\ref{g352.fig}, an exposure map and diffuse X-ray emission as part of a composite X-ray/IR image. There is a notable absence of diffuse emission along the IRDC filament, indicating strong absorption of background X-ray emission by the cold, dense molecular cloud.
    
%  \todo{Pat --- do you have any  notes about any data processing that went beyond what was already described in the MOXC papers? \emph{AE v. 5551 2020-07-25} If not, that's fine we can just cite previous work.}
    
An electronic table containing all 11,470 point sources extracted from the 3-degree mosaic accompanies this paper (\texttt{xray\_properties\_G352-panoramic.fits}). If using this table, please also cite \citet{MOXC2,MOXC3}, the original sources for the NGC 6334 and NGC 6357 MOXC catalogs, respectively.  
    %Matt made this with 14,470 sources in TOPCAT (see TOPCAT\_analysis file).  
    For descriptions of the standard MOXC data analysis procedures and the resulting columns of this table, please see \citet[][their Table 3]{MOXC3}.

%\input{observing_log}

% This is a combination of LKT's two observing log tables for this mosaic. Commented out the 100% discarded ObsIds.

%  2020 Aug 03 09:02
%  acis_extract, version 5551  2020-07-25; ae_flatten_collation, version 5547  2020-07-20; hmsfr_tables, version 5547  2020-07-20
% COLUMN SPACING
\setlength{\tabcolsep}{0.5mm}

% ROW SPACING 
% The traditional method for controlling row spacing, e.g.
%   \renewcommand{\arraystretch}{1.5} 
% does not work with AASTeX tables.  The AASTeX manual says you must edit the class file:
%
%    "Authors can control the space between data lines in tables by making a simple modification to the classfile. The line \def\arraystretch{1.0} is set to the default value of 1. To increase the space between data lines by 10% change the argument to 1.1. Likewise, to decrease the space between data lines by 10% to produce a tighter table use 0.9."

%\begin{longrotatetable}
%\startlongtable
\begin{deluxetable}{crrrccrlllllC}
\centering 
\tabletypesize{\tiny} \tablewidth{0pt}
%\tablecolumns{8}

\tablecaption{Log of {\em Chandra} Observations Newly Added to the MOXC Catalogs
 \label{tbl:obslog}}

\tablehead{
\colhead{Target} & 
\colhead{ObsID} & 
\colhead{Start Date} & 
\colhead{Exp} & 
\multicolumn{2}{c}{On-axis Position} & 
\colhead{Roll} & 
%\multicolumn{2}{c}{} & 
\colhead{Config} & 
\colhead{Mode} &
\colhead{PI}  &
\colhead{TGAIN} &
\colhead{OBF} &
\colhead{Shift}  \\
\cline{5-6}
\colhead{} & 
\colhead{} & 
\colhead{(UT)} & 
\colhead{(ks)} & 
\colhead{$\alpha_{\rm J2000}$} & 
\colhead{$\delta_{\rm J2000}$} & 
\colhead{(\arcdeg)} & 
\colhead{} &
\colhead{} &
\colhead{} &
\colhead{} &
\colhead{} &
\colhead{(SKY pixel)} 
}
\decimalcolnumbers
\startdata
        G352.279$+$0.808 & \dataset[ 18909]{  \doi{10.25574/18909}} &       2017-07-05 &   39553 & 17:22:40.02 & -34:59:12.2 &  293 &     I (0123) &  TE-VF & Leisa Townsley & \path{ 2017-05-02N6 } & N0011 & (+0.538, -0.566) \\ % 0\% exposure discarded for high background.
        G352.015$+$0.705 & \dataset[ 18908]{  \doi{10.25574/18908}} &       2017-10-11 &   36979 & 17:22:19.69 & -35:15:54.9 &  260 &   I (012367) &  TE-VF & Leisa Townsley & \path{ 2017-05-02N6 } & N0011 & (+1.069, +0.690) \\ % 0\% exposure discarded for high background.
        G352.557$+$0.744 & \dataset[ 18910]{  \doi{10.25574/18910}} &       2017-10-13 &   39446 & 17:23:40.47 & -34:47:49.1 &  260 &   I (012367) &  TE-VF & Leisa Townsley & \path{ 2017-05-02N6 } & N0011 & (-0.013, -0.200) \\ % 0\% exposure discarded for high background.
%        G352.841$+$0.720 & \dataset[ 19705]{  \doi{10.25574/19705}} &       2017-10-14 &   39532 & 17:24:32.87 & -34:34:34.0 &  260 &    I (01237) &  TE-VF & Gordon Garmire & \path{ 2017-05-02N6 } & N0011 & (+0.377, -0.141) \\ % 0\% exposure discarded for high background.
        %%        NGC 6334 REGION 1 & \dataset[  2574]{  \doi{10.25574/02574}} &       2002-08-31 &       39.65 & 17:20:53.44 & -35:47:19.3 &  269 &   I (012367) &   TE-F & Yuichiro Ezoe & \path{ 2002-08-01N8 } & N0013 & (+0.161, +0.094) \\ % 100\% exposure discarded for high background.
%%         G352.015$+$0.705 & \dataset[ 18908]{  \doi{10.25574/18908}} &       2017-10-11 &       36.98 & 17:22:19.69 & -35:15:54.9 &  260 &   I (012367) &  TE-VF & Leisa Townsley & \path{ 2017-05-02N6 } & N0013 & (+1.069, +0.690) \\ % 100\% exposure discarded for high background.
         G351.744$+$0.594 & \dataset[ 21679]{  \doi{10.25574/21679}} &       2019-06-07 &   37.09 & 17:22:02.01 & -35:32:19.8 &   20 &     I (0123) &  TE-VF & Gordon Garmire & \path{ 2019-05-02N8 } & N0013 & (-0.591, -1.115) \\ % 0\% exposure discarded for high background.
         G351.613$+$0.847 & \dataset[ 21680]{        \doi{10.25574/21680}} &       2020-07-01 &   19.70 & 17:20:37.28 & -35:30:37.0 &  300 &   I (012367) &  TE-VF & Gordon Garmire & \path{ 2019-11-01_biN2_revA } & N0013 & (-0.102, +0.441) \\ % 0\% exposure discarded for high background.
         G351.613$+$0.847 & \dataset[ 23294]{        \doi{10.25574/23294}} &       2020-07-01 &   19.79 & 17:20:37.32 & -35:30:37.8 &  300 &    I (01237) &  TE-VF & Gordon Garmire & \path{ 2019-11-01_biN2_revA } & N0013 & (+0.263, +0.320) \\ % 0\% exposure discarded for high background.
\enddata
\tablecomments{
%The table is divided by target; the name of each target from Table~\ref{targets.tbl} is shown in bold.
Col.\ (1): Name of the target in the \anchorparen{https://cda.harvard.edu/chaser/mainEntry.do}{\Chandra Data Archive search and retrieval tool {\em ChaSeR}}.
\\Col.\ (2): \Chandra Observation Identification (ObsID) number.
\\Col.\ (3): Calendar date when the observation began.
\\Col.\ (4): Exposure times are the net usable times after various filtering steps are applied in the data reduction process. 
%For the following ObsIDs, we discarded exposure time as noted to remove periods of high instrumental background: 
%LIST HERE PERCENTAGE OF CCD FRAMES DISCARDED FOR EACH OBSID, e.g.
%3501 (2\%),
%6410 (3\%).
The time variability of the ACIS background is discussed in \S6.16.3 of the \anchorparen{http://asc.harvard.edu/proposer/POG/}{\Chandra Proposers' Observatory Guide} 
and in the ACIS Background Memos at \url{http://asc.harvard.edu/cal/Acis/Cal_prods/bkgrnd/current/}.
\\Col.\ (5) \& (6): The On-axis Position is the \anchorparen{http://cxc.harvard.edu/ciao/faq/nomtargpnt.html}{time-averaged location of the optical axis (CIAO parameters RA\_PNT,DEC\_PNT)}.
Units of right ascension ($\alpha$) are hours, minutes, and seconds; units of declination ($\delta$) are degrees, arcminutes, and arcseconds.
\\Col.\ (7): Observatory roll angle.
\\Col.\ (8): ACIS observations can take on three \Chandra configurations: I (optical axis on ACIS-I, no grating), S (optical axis on ACIS-S, no grating), or H (optical axis on ACIS-S, HETG in place).
The subset of the ACIS CCD detectors enabled for the observation is listed in parentheses; the layout of the ten detectors (numbered 0--9 here) in the ACIS focal plane is shown in \S6.1 of the \anchorparen{http://asc.harvard.edu/proposer/POG/}{\Chandra Proposers' Observatory Guide}.
\\Col.\ (9): ACIS observing modes are described in \S6.12 of the \anchorparen{http://asc.harvard.edu/proposer/POG/}{\Chandra Proposers' Observatory Guide}.
\\Col.\ (10): Principal Investigator of the observation.
\\Col.\ (11): Abbreviated name of the ACIS Time-Dependent Gain file used for calibration of event energies, e.g., ``2002-02-01N6'' = ``acisD2002-02-01t\_gainN0006.fits''.
\\Col.\ (12): The version of the Optical Blocking Filter model used for calibration of Ancillary Response Files and exposure maps.
\\Col.\ (13): The shift (in RA and Dec) applied to the ObsID's aspect file (via the {\em CIAO} tool {\em wcs\_update}) to achieve astrometric alignment, 
expressed as (dx,dy) in the \anchorparen{http://asc.harvard.edu/ciao/ahelp/coords.html}{\Chandra ``SKY'' coordinate system}; 1 SKY pixel = 0.492$\arcsec$.
}
\end{deluxetable}
%\end{longrotatetable} 

\bibliography{citations}{}
\bibliographystyle{aasjournal}



\end{document}